# A Magnetotelluric Instrument for Probing the Interiors of Europa and Other Worlds

Robert Grimm[1], Ton Nguyen[2], Steve Persyn[2], Mark Phillips[2], David Stillman[1], Tim Taylor[2], Greg Delory[3], Paul Turin[3], Jared Espley[4], Jacob Gruesbeck[4], Dave Sheppard[4]

[1]Southwest Research Institute, Boulder, Colorado, USA

[2]Southwest Research Institute, San Antonio, Texas, USA

[3]Heliospace Corp., Berkeley, California, USA

[4]NASA Goddard Spaceflight Center, Greenbelt, Maryland, USA

Corresponding author: Robert Grimm, grimm@boulder.swri.edu

**Abstract**

One objective of a lander mission to Jupiter's icy moon Europa is to detect liquid water within 30 km as well as characterizing the subsurface ocean. In order to satisfy this objective, water within the ice shell must also be identified. Inductive electromagnetic (EM) methods are optimal for water detection on Europa because even a small fraction of dissolved salts will make water orders of magnitude more electrically conductive than the ice shell. Compared to induction studies by the Galileo spacecraft, measurements of higher-frequency ambient EM fields are necessary to resolve the shallower depths of intrashell water. Although these fields have been mostly characterized by prior missions, their unknown source structures and plasma properties do not allow EM sounding using a single surface magnetometer or the orbit-to-surface magnetic transfer function, respectively. Instead, broadband EM sounding can be accomplished from a single surface station using the magnetotelluric (MT) method, which measures horizontal electric fields as well as the three-component magnetic field. We have developed a prototype Europa Magnetotelluric Sounder (EMS) to meet the measurement requirements in the relevant thermal, vacuum, and radiation environment. EMS comprises central electronics, a fluxgate magnetometer on a mast, and three ballistically deployed electrodes to measure differences in surface electric potential. In this paper, we describe EMS development and testing as well as providing supporting information on the concept of operations and calculations on water detectability. EMS can uniquely determine the occurrence of intrashell water on Europa, providing important constraints on habitability.

1. Introduction

Due to its subsurface ocean, Jupiter's moon Europa is a potentially habitable world. It is vigorously resurfaced (e.g., Figueredo and Greeley, 2004), suggesting that recently-erupted material could yield key information about interior habitability or even direct indicators of life. NASA's most recent analysis of a surface mission (Europa Lander Study 2016 Report: Hand et al., 2016) therefore focused on searching for biosignatures and characterizing their surface context. However, an important corollary investigation was to determine the relationship of the landing site to any liquid water, i.e., the subsurface ocean or regions within the ice shell. We began developing a Europa Magnetotelluric Sounder (EMS) around the time this study was completed and EMS was subsequently selected for further maturation in collaboration with the mission pre-project. This paper first gives the rationale for a magnetotelluric (MT) instrument—including comparison with seismology, surface-penetrating radar, and prior induction studies in the jovian system—to achieve the objective of intrashell water characterization on Europa. This approach is supported with detailed 1D calculations of a variety of possible subsurface structures and we also show that limited 3D information can be obtained from a single station. Functional and deployment testing is illustrated for the first completed ESS prototype, as well as work underway for a second prototype with improved performance. The latter is the basis for the Lunar Magnetotelluric Sounder (LMS) selected for flight to the Moon in 2023. We conclude with a brief overview of additional applications of MT experiments elsewhere in the Solar System.

## 2. Water Detection on Europa

Hand et al. (2016) posed the objective: "Search for any subsurface liquid water within 30 km of the lander, including the ocean" (**Fig. 1**). A Geophysical Sounding System (GSS) was named to achieve this objective and a seismometer (baseline) or geophone (threshold) mounted within the spacecraft was given as the GSS model payload. However, the ability of seismology to answer this specific question is uncertain. First, many events (quakes) are necessary to construct traveltime curves (e.g., Fig. 11 in Pappalardo et al., 2013). These events must be distributed across all distances within 100 km or so of the lander and of course must occur within the nominal lander lifetime of a few weeks. Cyclic events on the same fault add no variety. If these criteria are met, the depth to the ice-ocean interface can be determined from its principal reflection. Pappalardo et al. (2013) then invoke azimuthal anisotropy to detect lateral anomalies that may be due to water layers. This requires an order of magnitude more events and assumes wide source variety in both azimuth and range, so that small spreads in traveltime at each source distance can be translated to lateral heterogeneity. While we support a seismometer on a Europa lander, the present concept for detection of intrashell water has such large uncertainties that it is prudent to consider alternative methods.

An ice-penetrating radar (IPR) provides excellent structural subsurface mapping in terrestrial glacial environments (e.g., *MacGregor* et al., 2015). IPRs will fly on both the ESA JUICE and NASA Europa Clipper missions in the 2020s (Bruzzone et al., 2013; Blankenship et al., 2018). However, electromagnetic energy is absorbed with increasing ice temperature and soluble impurities, namely chloride and acid (e.g., Stillman et al., 2013). If a 10-km thick convecting ice shell contains just 30 µM lattice chloride, reflections cannot be obtained deeper than ~4 km, given the maximum 60-dB dynamic range available in the ground for IPR (Stillman et al., 2018). These arguments can be extended to intrashell water with overlying conductive thermal profiles: there is insufficient dynamic range to record reflections from water layers > 2-3 km deep. Fading of reflections within the overlying ice may indicate absorption caused by the warm water layer, but direct detection of intrashell water can be achieved only within a small fraction of the 30-km distance set out to understand the geological context of the lander. Furthermore, radar signals will not be able to penetrate through brackish or briny intrashell water to characterize thickness or salinity.

Detection of liquid water in Europa is performed optimally and with the least risk using inductive electromagnetic (EM) methods, from two principal considerations. First, water with even a small quantity of dissolved salts is a strong conductor compared to surrounding ice: finding conductors is the essential ability of inductive EM (e.g., the familiar metal detector). Second, the EM fields around Europa that will serve as sources are already known to exist. The ambient energy available is enormous compared to radar, such that loss tangent $\gg$ 1 is sustainable. Together, these considerations also imply that information about the conductor's thickness and electrical conductivity (salinity) can also be gleaned.

## 3. Electromagnetic Sounding

*3.1 Electromagnetic Principles*

A time-varying magnetic field causes eddy currents to flow in a target with finite electrical conductivity. These currents in turn produce a secondary magnetic field that opposes the primary, such that the net magnetic field decays with 1/e folding ("skin") depth $\sqrt{2/\mu\omega\sigma}$, where $\mu$ is the permeability, $\omega$ is the angular frequency, and $\sigma$ is the conductivity. The skin-depth effect is the heart of EM sounding that allows frequency-dependent measurements of EM fields to be translated into conductivity as a function of depth.

Although it is possible to fit EM fields directly, solid-earth geophysics has preferred to translate these data to proxies that have physical meaning. Here we focus on apparent conductivity $\sigma_a$, which is the conductivity of a half-space having the same response as the target. Then EM sounding can be considered in terms of using the skin-depth effect to invert $\sigma(z)$ from $\sigma_a(f)$, where $z$ is depth and $f$ is frequency. Alternatively, the (complex) impedance $Z$ ($\sigma_a = \mu\omega/|Z|^2$) can be the most descriptive when considering measurement methods: the familiar Ohm's Law $Z = V/I$ calls for two independent quantities to determine the impedance, and the same applies to EM sounding.

There are two approaches to determining $\sigma_a$ or $Z$ that are relevant to Europa (see Grimm and Delory, 2012; Banerdt et al., 2014). The Transfer Function (TF) compares the magnetic field at or near a planetary body (the sum of source and induced fields) to a known, distant source field. Direct measurement of the source field by the Explorer 35 satellite set up the transfer function to the Apollo 12 lunar surface magnetometer (e.g., Sonett, 1982). The Galileo induction studies can be considered a special case of the TF, wherein very low-frequency source fields are known and modeled a priori (e.g., Khurana, 1997). It is important to recognize that if the source field is not known independently, a single magnetometer is insufficient to perform EM sounding. This also rules out Geomagnetic Depth Sounding approaches commonly used for Earth.

In the Magnetotelluric (MT) method, the electric field $E$ and the magnetic field $B$ are used to form $Z = \mu E/B$, where $E$ and $B$ are in fact measured orthogonally and parallel to the surface. Because $E$ supplies the required second piece of information, the source field does not need to be known. Tensor impedance measurements (using both horizontal components of both $E$ and $B$) further enable anisotropy, or directionality of targets, to be determined (see below). MT is the most highly developed and widely applied EM method for investigations of Earth's crust and upper mantle (e.g., Vozoff, 1991; Simpson and Bahr, 2005; Chave and Jones, 2012).

*3.2 EM Sources at Europa*

Jupiter's magnetic dipole axis is inclined with respect to the orbital plane of the Galilean satellites, so rotation of Jupiter's magnetosphere creates a time-varying field at the synodic period of 11.23 hrs. There is also a smaller contribution from rotation of the current sheet and inclination and eccentricity of Europa's orbit. The large signal at the synodic period was first used to detect Europa's subsurface ocean (Khurana et al., 1998). Seufert et al. (2011) described the possible contributions of the quadrupole and octupole of the synodic period (higher frequency), the inclination and eccentricity of Europa's orbit (low frequency) and magnetopause variability due to the solar rotation (the longest period). We call these kinematic signals because they arise from relative motion (**Table 1**). These signals are expected to be as small as ~1 nT, which will be extremely challenging for orbital measurements and are best made from a lander. Because the kinematic sources are known independently, induction responses at periods of 3 to 700 hr can be analyzed using a magnetometer alone (special case of TF method).

Magnetospheric signals span a broad range of frequencies from less than 0.01 Hz to greater than 1 kHz, all orders of magnitude higher than considered heretofore for Europa EM sounding. Due to their smaller skin depths, these frequencies play the pivotal role in constraining the properties of the ice shell (see below). Near Europa, magnetospheric signals may arise from the Kelvin-Helmholtz instability, the drift-mirror instability, ion-cyclotron instabilities, or whistlers (Khurana and Kivelson, 1989; Wilson & Dougherty, 2000; Glassmeier et al., 1989, Gurnett et al., 1998; Kurth et al., 2001; Volwerk et al., 2001). Amplitudes are again in the nT range.

Because the strength and geometry of incident magnetospheric waves are not known independently, the surface magnetic-field measurement must be supplemented by an orbital magnetometer or a surface electric-field measurement. The orbit-to-surface transfer function was implemented for Apollo-Explorer up to ~30 mHz, but suffered from plasma distortion above several mHz (Sonett et al., 1972). Here the wavelengths of the source magnetohydrodynamic waves become comparable to the circumference of the Moon, which causes higher harmonics (multipoles) to be excited. This requires joint treatment of the plasma properties and the lunar interior, increasing uncertainty. In contrast, the MT method is largely insensitive to source-field properties (Grimm and Delory, 2012). Furthermore, broadband MT data allow any remaining noninductive signals to be identified and removed.

## 4. Modeling

We demonstrate the power of EM sounding for water characterization on Europa with a suite of 1D models for the induction response of a layered Europa interior (**Fig. 2**) and examples of 3D calculations to discriminate target shape (**Fig. 3**). The 1D computations translating $\sigma(z)$ to $\sigma_a(f)$ are described by Grimm and Delory (2012). They use a plane-layer propagator (Wait, 1970) with spherical transformation (Wiedelt, 1972). This allows a finely discretized, arbitrary radial conductivity structure.

The reference model consists of an ice shell with an interior water layer, an ocean, a silicate mantle, and a metal core. The intrashell water-layer depth, thickness, and conductivity are 3 km, 2 km, and 1 S/m, respectively; ice-shell thickness and conductivity are 10 km and $10^{-4}$ S/m, respectively; ocean thickness and conductivity are 100 km and 1 S/m, respectively, and mantle conductivity is $10^{-3}$ S/m. Rather than perform formal and well-established inversions for $\sigma_a(f)$ to an estimated conductivity profile $\sigma_{est}(z)$, we highlight the diagnostic parts of these curves that control the inversions. Horizontal segments indicate a range of frequencies that penetrate into a layer but not through it, thus $\sigma_{est} \approx \sigma_a$ there. Slopes approaching $-45°$ yield the depth to a conductor but not its bulk properties, as these frequencies are not sufficiently penetrating the target. Slopes approaching +45° indicate frequencies penetrating through a conductor and yield only the conductivity-thickness product (the conductance). When all three of these behaviors are evident for a conducting target layer, its depth, conductivity, and thickness can be uniquely determined. This is a consequence of broadband data and the skin-depth effect. Nonuniqueness is introduced according to which segment(s) are missing.

Ocean thickness and conductivity both strongly affect $\sigma_a$ at the kinematic frequencies (**Fig. 2**). Where these frequencies straddle the local maximum, separation is possible—this is why Europa Clipper will attempt to detect two or more kinematic frequencies during its flybys. The response plots in Khurana et al. (2009) and Seufert et al. (2011) highlight that unique solutions using the synodic and orbital exist only in certain portions of the conductivity-thickness space. In the magnetosphere band, there is no response to different thickness and conductivity, because these signals negligibly penetrate the ocean. Both the kinematic and magnetosphere signals have some sensitivity to ice-shell thickness, with opposite, complementary, dependencies.

In contrast, the kinematic frequencies have no sensitivity to the depth or thickness of a water layer in the ice shell at nominal conductivity and only begin to respond when conductivity is very high. However, clear responses are evident in the magnetospheric band for all of these properties; in fact, diagnostic responses are mostly bounded between 0.01 and 100 Hz. The depth to a water layer is always well-determined. There is some ambiguity between conductivity and thickness in perturbations of the reference model, which is eliminated for water layers that are either shallower, thinner, or less conductive, or for a thicker ice shell.

The 3D calculations (**Fig. 3**) were performed using Comsol Multiphysics®. The ocean conductivity was increased to 10 S/m while the water-in-shell conductivity was decreased to 0.1 S/m and the layer thickness reduced to 1 km, all to evaluate more challenging discrimination. This sill-like geometry was compared to a vertical layer (dike) with the same thickness and 3-km lateral offset from the lander, and to a 3-km radius sphere offset by 3 km both laterally and vertically. Apparent conductivities were normalized by the ocean response due to the dominant influence of the latter. The responses of the dike and sphere are anisotropic; rotating the impedance tensor yields the direction to the target and the principal modes, transverse electric (TE) and transverse magnetic (TM). In this example, the effect of the target is maximized for the former. Exploiting anisotropy, each of these three canonical targets can be distinguished from each other and from the ocean.

## 5. Instrument

EMS is a new system designed to apply terrestrial MT to other worlds, but its subsystems have significant spaceflight heritage as described in individual sections below. EMS is, however, much more compact than commercial MT systems: the latter use large searchcoil magnetometers and ground-coupled electrodes typically separated by 100 m (**Fig. 4**). EMS uses a small fluxgate magnetometer and plasma-coupled electrodes on a minimum baseline of 20 m (**Figs. 5, 6**). Signal-to-Noise (SNR) analysis (**Table 2**) demonstrates that the fluxgate magnetometer sensitivity has raw SNR >5 throughout the target band 0.01 to 100 Hz, whereas the electrometer raw SNR is always >30 using a 20-m baseline. Root-mean-square (RMS) signals ~3 nT and ~3 mV are expected. Because the impedance is formed from *E*/*B*, the variances (squared reciprocals of SNR) are added to determine the joint SNR.

Extensive signal integration is possible because the instrument is never moved: in 16 hours, joint SNRs > 100 may be achieved. This is a small fraction of the ~120 hrs available for MT when the lander is in daylight and quiescent (Krajewski et al.,, 2018). EMS requires photoelectron coupling for the E-field measurement and so only operates in daylight. The risk of lander noise is minimal during quiescent periods and there is ample measurement time for both editing and stacking in order to perform robust measurements. Recall that the magnetometer alone in principle can be used to determine the impedance at the kinematic frequencies <$10^{-4}$ Hz, for which the source field is known.

We describe the Europa Magnetotelluric Sounder, first as individual subsystems (central electronics, magnetometer, electrometer, and deployment mechanisms) and then its integrated functional test and radiation mitigation. The first prototype (EMS-1) meets thermal-vacuum requirements for TRL 6 (prototype instrument in a relevant environment, defined in **Table 3**). The second prototype (EMS-2) is in progress; additional electrometer circuitry will improve those measurements in Europa's plasma environment. Radiation mitigation was assessed by theoretical calculation and would not be implemented until flight development. The principal organizations are Southwest Research Institute (SwRI: electronics, integration and test), Heliospace Corp.

(electrodes, electrometer-circuit design, deployment mechanisms) and Goddard Spaceflight Center (GSFC: magnetometer sensor and board).

*5.1 Central Electronics*

The EMS electronics comprise five individual boards: low-voltage power supply, digital board, analog board, floating voltage supply board, and magnetometer board (**Figs. 7, 8**). The last is a MAVEN engineering model provided by GSFC and described separately below. The SwRI electronics use Industrial Temp-Commercial and Military Engineering components to achieve an engineering-model level required to meet TRL 6. All these EEE components were chosen to have a direct path to Level 2 QML-Q flight components or better.

The Low-Voltage Power Supply (LVPS) receives primary power from the lander (unregulated at 24-34 Vdc), performs DC to DC conversion, and delivers secondary power to the remainder of the system. LVPS is build-to-print from the Radiation Assessment Detector (RAD) instrument on the Mars Science Laborarory (MSL: Hassler et al., 2012), with the exception of the added Frangibolt switches that activate the sensor deployments.

The Floating Voltage Supply Board (FVSB) provides signal conditioning necessary for the electric-field measurement system (electrometer) to function under Europa conditions, in which probe-to-plasma potentials could be as high as 75 V and thus saturate the response. To mitigate this possibility, a standard bias-current circuit and floating power supply is included (e.g., Bonnell et al., 2008). The bias current is supplied by a programmable voltage offset through a large resistance to the electrometer input. It is chosen to offset the corresponding plasma currents to the sensor and thus reduce the overall DC offset and effective coupling impedance between the electrometer and the plasma. The use of the bias current, and differences between the electrometer and lander electrical properties, could also result in large quasi-DC potentials between the lander and the sensors that could saturate the response. Therefore the electrometer power supply is bootstrapped to the electrometer output, which may drift up to ±90 V relative to lander ground. This enables measurements over the full dynamic range even if the lander power supply voltages are exceeded.

The Analog Board (AB) and Digital Board (DB) are also derived from RAD. The AB includes additional amplification and filtering of the electrometer signal and performs the analog-to-digital conversion at 256 Hz sampling rate. The middle electrode can serve as the reference and so two differential channels would be sufficient, but the potential of the reference electrode to the lander is also recorded. The DB receives, stores, and transfers digital data from both the electrometer and magnetometer. It also commands the sensors and controls the interaction with the lander communication interfaces. The software in the 8051 microcontroller for both instrument commanding and science data analysis is based on RAD architecture, with minor modifications to accommodate the instrument-unique science data processing. The latter is simply anti-aliased downsampling and 50% lossless compression. This limits data volume by returning data at lower rates that still determine the responses of lower parts of the target frequency band. More complex onboard processing (e.g., FFT or cascade-decimation spectral estimation) were considered but set aside for now because the current approach returns complete time-series data and satisfies the science objectives.

*5.2 Magnetometer*

The magnetometer subsystem (MAG) is provided by the NASA Goddard Spaceflight Center (GSFC). The three-component fluxgate sensor and the electronics board are both engineering models from the MAVEN mission (Connerney et al., 2015a). For flight, the new GSFC "miniMAG"—an update of the sensor used for the Polar Orbiting Geomagnetic Survey (POGS) mission—would be used. Two 16-bit measurement ranges (±65,536 nT high and ±1024 nT low) are available, but automatic ranging will always select the latter at Europa, where the maximum quasi-static field of 460 nT will at most double in the presence of a perfect conductor. This gives A/D conversion precision of 31 pT.

MAG operates with the application of power; it requires no configuration commands. MAG has no processor and no software; instead all logical functions are performed via finite state machines contained in the radiation-hardened (300 krad) FPGA. The electronics drive the fluxgate ringcore at a frequency synchronized to the master oscillator to preclude beat-frequency interference. As in previous missions, the sensor utilizes lightweight composites. The sensor connection to the electronics boards is accomplished via a resonantly tuned circuit, which is calibrated at GSFC. The sensor absolute vector accuracy is ±1 nT + 1% of |B|. The offset drift is <1 nT/yr (Connerney et al., 2015b). Note that drift has negligible impact on our investigation, as its rate is much slower than any frequency of interest. The sensor alignment accuracy is ±0.5 degrees with an alignment knowledge of <0.1°.

Fluxgate sensors are commonly blanketed and heated to a calibrated temperature range at a cost of a watt or more. We wanted to assess the effects of an unheated sensor (**Fig. 9**) on performance, in order to save power on a severely energy-limited (battery-powered) mission. Using the unique non-magnetic thermal-vacuum chamber at the GSFC Mag Lab, we have evaluated standard magnetometer-performance criteria (orthogonality, scale factor, and offset) at temperatures 80 to 290 K.

The first major test result was that the sensor functioned without any difficulties at 80 K: this is a first for fluxgate magnetometers. In detail, the orthogonality was generally within 1‰ but the XZ variation was as high as ~3‰ at 125–150 K (**Fig. 10a**). Scale factors changed up to 1% over the entire temperature range but are very linear and so can be predicted to <1‰ (Fig 10b). Static offsets did vary in a significant and nonlinear way with temperature by as much as 20 nT. Because the DC level is subtracted prior to spectral analysis, this behavior is irrelevant to the EMS investigation. Therefore the XZ orthogonality deviation can be taken as representative of the maximum error, which translates to 15 pT at a typical source amplitude of 3 nT. This is less than the A/D precision and so can be neglected. We can conclude that a fluxgate magnetometer sensor can return accurate measurements even while unheated at the surface of Europa.

*5.3 Electrometer*

The electrometer system comprises the three remote electrodes (REs), the FVSB in the central electronics, and 20 m of coaxial cable connecting each RE to the FVSB. It is based on similar instruments flown on THEMIS and other missions (e.g., Bonnell et al., 2008). The FVSB—whose purpose is to minimize probe-to-plasma voltage offsets that could saturate the ADC range—was described above. The RE itself is essentially a metal shell connected to a preamplifier (**Fig. 11**). The shell provides a large area for electrical coupling as well as mass for ballistic deployment. The preamp has been assembled and tested for input impedance, leakage/stray currents, and performance over the expected bandwidth. Input resistance is $>10^{12}$ ohms at DC and leakage currents <200 fA, which give high confidence that accurate potential measurements can be made without disturbing the ambient plasma. The outer shell coating can be vitreous carbon (DAG) as on previous designs, or inert metallic (i.e., Au coated) for durability.

*5.4 Deployment*

The magnetometer must be stood off from the lander in order to minimize any extraneous signal. Our new mast design uses two Ametek Stacers to displace the magnetometer 2.5 m but accurately control its final orientation (**Fig. 12**). Each Stacer is a self-extending, helically wound strip of hardened beryllium copper that packs into a small canister. Deployment is actuated using an Ensign-Bickford Aerospace Frangibolt, which is a shape-memory-alloy based release mechanism. As the Stacer coils extend, they rotate and collapse into a helix forming a rigid tapering tube. A control lanyard absorbs residual energy at end of the 2.5-m extension. An electronic harness connects the magnetometer to the base, and thence to the central electronics. The mast is sized for the miniMAG and therefore cannot directly accommodate the larger MAVEN sensor.

The electrodes must be remotely deployed to produce enough physical separation to read out a robust voltage difference. In addition to physical separation, the electrodes must be in ground contact, resting on the "electrical plane" of the surrounding environment with measurements uncontaminated by any vertical component of the electric field. This placement of electrodes on a planetary surface and immersed in a conductive plasma is analogous to contemporary seafloor MT, with electrodes immersed in conductive seawater. We developed a ballistic launcher (**Fig. 13**) in order to achieve a ground-contacting electrode at a distance much larger than possible with a boom. The launcher is spring powered and is released by a Frangibolt. The ejected electrode unwinds the comm/power cable. A heater is included to bring the cable to a suitable temperature for deployment. Our nominal design uses three co-located electrodes (Fig. 5) deployed at 90°-separations to 14-m distance. In this way, two orthogonal 20-m baselines are formed (Fig. 6).

To verify mast and electrode deployment system performance at temperatures typical of the surface of Europa, deployment tests were conducted under thermal vacuum (TVAC) conditions. A TVAC chamber fitted with an internal thermally isolated box connected to two cryocoolers enabled test temperatures of 80 K or lower at pressures $<10^{-3}$ Pa. A single electrode launcher was mounted on one end of the box, and a basket made of Multilayer Insulation (MLI) secured at the opposite end to catch the electrode to avoid any component or chamber damage during deployment. Electrode launch was successful at 80 K (**Fig. 14**), which validates both the Frangibolt deployment mechanism and verifies that no binding occurred as the electrode and cable left the launcher into its free trajectory to the opposite side of the test box. Simultaneously with the electrode deployment test, a separate electrode preamp board was tested for input and gain characteristics typical of local plasma conditions over the same temperature range.

Deployment tests of the magnetometer mast were conducted in the same test apparatus using a horizontal deployment track. These tests used a sub-scale 1-m length version of the mast and the identical deployment mechanism planned for flight. Horizontal deployments typically represent an over-test as the self-extracting Stacers overcome friction along the track. Initial tests indicated that tolerance stack-up issues in the Frangibolt release mechanism became important below 190 K; after proper shimming, successful deployments were achieved at 80 K (Fig. 14).

*5.5. Functional Testing*

In practice, EMS comprises two prototypes according to the funding sequence. EMS-1, completed as of this writing, includes the MAVEN engineering-model fluxgate magnetometer and digital board, two remote electrodes, and digital, analog, and low-voltage power supply boards. This version was functionally tested under (near) Europa conditions, as described below. EMS-2 is in progress and is responsive to the specific requirements of the Europa lander as set out by Krajewski et al. (2018). It adds a third remote electrode and the analog circuitry (bias current and floating-voltage power supply) necessary to adjust to changing plasma conditions. The subsystems described above are the EMS-2 configuration, which would be subsequently proposed for flight.

Functional testing of EMS-1 was performed in TVAC with a configuration representing both the interior and exterior of the lander (**Fig. 15**). The 8 x 8 ft SwRI chamber was configured with the platen (lander deck) at 123 K and the shroud (environment) at 93 K. The electronics were maintained at 228 K by thermal blanketing, heating, and elevation from the platen. When operating, the electrodes were stable around 137 K and the magnetometer was close to ambient temperature. The magnetometer recorded ambient fields and the electrodes were grounded through a 1 MΩ resistor (a simulated contact resistance). Differences in magnetic fields between room temperature (291 K) and near-Europa conditions (**Fig. 16**) were small, consistent with the separate low-temperature testing described above. The strong powerline and vibrational interference emphasizes that this was a functional test and not a calibration. The electrometer signals differed more but were still comparable (Fig. 16). This difference was dominantly at the higher frequencies.

We also tested the offset between the electric and magnetic signals. Close agreement is needed because phase is also important to the MT analysis. At room temperature, a 1-Hz signal was applied across the electrodes using a resistively loaded wire and to the fluxgate sensor using a wrapped coil in a triple mu-metal shield. The cross correlation was fit to the expected cosine function and interpolated at 10 μs. The minimum error between the theoretical and fit functions is at −0.28 ms (**Fig. 17**) is a small fraction of the 3.9-ms sampling rate.

*5.6. Radiation Mitigation*

Due to the severe radiation environment, The Europa Lander accommodates avionics and science instruments in a shielded vault (Krajewski et al., 2018). The Total Ionizing Dose (TID) within the vault is expected to be 150 krad (Si equivalent) and a Radiation Design Factor (RDF) = 2 is required. The flight electronics for EMS are rated to 75-100 krad. Therefore some combination of a secondary enclosure and spot-shielding must be used (e.g., McComas et al., 2017). For preliminary design purposes, the EMS electronics can be enclosed in a 15-mil thick titanium box and thus mitigate the specified 300 krad to 88 krad. In practice, a composite material would be used to reduce the production of secondaries. Tantalum spot shielding can then bring sensitive components within rating. The mass of the secondary enclosure is accounted for in the EMS resource requirements.

Outside the vault, TID 1.7 Mrad (behind 100 mil, Si equivalent) is expected (Krajewski et al., 2018). The fluxgate sensor and deployment mechanisms have no active electronics and therefore are unaffected at the specified radiation levels. However, each electrode contains an operational amplifier in a TO-99 container that must be protected. We used Geant4 ray tracing to assess the dose as a function of different tantalum shield configurations. We found that a satisfactory full-shield design would not fit within the existing electrode volume and would unbalance the electrode, adversely affecting ballistic flight. Instead, a split-shield design was adopted that uses two separate pieces around the PCB (**Fig. 18**). The bottom piece has larger diameter to prevent leakage. For 3.5 mm Ta thickness, the computed dose is 82 krad (Si). The shielding adds ~40 g to the 270-g electrode.

## 6. Operations Concepts for Europa and Other Worlds

The battery-powered Europa Lander is strongly constrained to execute its surface science mission during daylight in under three weeks (Krajewski et al., 2018). Even then, the lander is active only at Europa Local Solar Time (ELST) approximately 7-10 hr and 14-17 hr. The lander is asleep the rest of the day and night, and allows undisturbed geophysical monitoring. We selected the broadest range of lander-sleep daylight hours for EMS operations (**Table 5**), with data acquisition intervals long enough to achieve adequate SNR at the lowest frequencies (Table 2). The adopted timeline specifies data acquisition after dawn, in late morning, in early afternoon, and before dusk. Each interval is 120 min long and uses three different sampling rates to cover the target bandwidth. The orbital period can also be fit to the magnetometer data using this scheme (adjustments of the individual window lengths would be required to sample the synodic periods). Acquisition over two Europa days allows repeat observations of these positions in Europa's orbit. EMS would consume 7% of the 1600 W-hr instrument-energy allocation and 14% of the 600 Mbit instrument-data allocation.

EM sounding can be efficiently applied at other worlds, and MT offers broadband, plasma-insensitive, single-station soundings elsewhere as well. Although any planet or moon with time-varying EM fields due to solar wind, magnetosphere, ionosphere, or lightning can be considered, we highlight the Moon, Mars, and Ceres due to their accessibility and investigations that are clearly posed at present.

Improvements to understanding vertical differentiation and lateral heterogeneity of the Moon afforded by renewed EM sounding were reviewed by Grimm and Delory (2012). In particular, MT is able to treat frequencies >> 10 mHz from solar-wind and magnetosphere sources that were obscured by finite plasma wavelengths in the Apollo-Explorer transfer functions. A Lunar Magnetotelluric Sounder (LMS), derived from EMS-2, has been selected for flight under the NASA Commercial Lunar Services Program (CLPS). A site in Mare Crisium has been approved, which will provide a measure of the background interior temperature and composition separate from the anomalous Procellarum KREEP Terrane that dominated Apollo investigations. MT is a core instrument of the proposed Lunar Geophysical Network (Shearer, 2011, Neal et al., 2020) at the New Frontiers level.

Groundwater on Mars is likely saline and therefore would be an ideal EM sounding target (Grimm, 2002): However, there has been no confirmation of natural signals at frequencies >0.1 Hz (e.g., lightning) required for probing groundwater at depths of several km using the MT method. Given this present uncertainty, active-resource methods (using a transmitter) at higher resource levels are favored for groundwater detection (Grimm et al., 2009; Stamenković et al., 2020). However, magnetospheric waves observed at 2-20 mHz (Chi et al., 2019) could be used for sounding of the martian upper mantle using MT, in good analogy with terrestrial studies.

A global subsurface brine layer at ~40 km depth on Ceres (Castillo-Rogez et al., 2020) would likewise be a very attractive EM sounding target. In this case, as with the Moon, the solar wind can provide a useful source. Both the global brine layer and possible intracrustal water are analogous in depth scales to Europa and are be treated in detail by Grimm et al. (2020).

## 7. Conclusion

A Europa Lander would provide a vital in situ assessment of the habitability of the archetypical "ocean world." EM sounding can not only determine the thickness and salinity of that subsurface ocean, but can also detect water bodies within the icy shell. The necessary source fields have already been partly characterized. The magnetotelluric method is the best approach to intrashell water detection: by measuring both electric and magnetic fields at the surface, MT can capture the higher frequencies required for shallower investigation without needing a reference magnetic field or being subject to plasma distortion. Our Europa Magnetotelluric Sounder is a medium-to-high fidelity prototype system, whose sensors, deployment mechanisms, and electronics have been tested at or near Europa surface temperatures. MT is broadly applicable to sounding of planetary interiors and is suitable for stand-alone investigation or as complementing seismology, geodesy, and especially heat-flow measurements.

## Acknowledgements

This work was funded by NASA grants 80NSSC17K0347 (COLDTech) and 80NSSC19K0609 (ICEE-2). We thank Ron Black for assistance with the ICEE-2 proposal and Roberto Monreal for the electrode radiation-shielding design.

**Tables**

**Table 1**. EM Source Frequencies and Strengths at Europa

| Period, hr | Frequency, Hz | Source | Amplitude, nT |
|---|---|---|---|
| 641.9[a] | 4.3x10$^{-7}$ | Magnetopause Variability from Solar Rotation | 1.1 |
| 85.22[a] | 3.3 x10$^{-6}$ | Orbital: Inclination, Eccentricity | 19 |
| 11.23[a] | 2.5 x10$^{-5}$ | Synodic: Dipole, Current Sheet | 200 |
| 5.62[a] | 4.9 x10$^{-5}$ | Synodic: Dipole, Quadrupole | 26 |
| 3.74[a] | 7.4 x10$^{-5}$ | Synodic: Quadrupole, Octupole | 2.3 |
| 10$^{-3}$ s–10$^{3}$ s[b] | 10$^{-3}$ to 10$^{3}$ | Magnetospheric Waves | 1–10 |

[a]*Seufert et al.* (2011). Mixed multipoles at same period are due to equator crossing in addition to azimuthal variability.

[b]*Khurana and Kivelson*, 1989; *Wilson & Dougherty*, 2000; *Glassmeier et al.*, 1989; *Kurth et al.*, 2001; *Volwerk et al.*, 2001.

**Table 2**. EMS Signal-to-Noise (SNR)

| Frequency, Hz | 0.01 | 0.1 | 1 | 10 | 100 |
|---|---|---|---|---|---|
| [a]B Signal, nT/√Hz | 45 | 4.5 | 0.45 | $4.5 \times 10^{-2}$ | $4.5 \times 10^{-3}$ |
| B Noise, nT/√Hz | $9 \times 10^{-4}$ | $9 \times 10^{-4}$ | $9 \times 10^{-4}$ | $9 \times 10^{-4}$ | $9 \times 10^{-4}$ |
| B Raw SNR (ampl.) | $5 \times 10^{4}$ | $5 \times 10^{3}$ | 500 | 50 | 5 |
| [b]App. Conductivity, S/m | 0.23 | $8.8 \times 10^{-2}$ | $1.2 \times 10^{-2}$ | $1.3 \times 10^{-3}$ | $1.4 \times 10^{-4}$ |
| [c]E Signal, μV/m/√Hz | 21 | 11 | 9.2 | 8.8 | 8.5 |
| [d]E Noise, μV/m/√Hz | 0.22 | 0.22 | 0.22 | 0.22 | 0.22 |
| E Raw SNR (ampl.) | 95 | 48 | 41 | 40 | 38 |
| [e]Joint Raw SNR | 95 | 48 | 41 | 31 | 5 |
| [f]Joint Stacked SNR | 140 | 230 | 410 | 980 | 210 |

[a]Regression of component mean amplitudes Galileo Europa encounters, extrapolated >1.5 Hz.

[b]Baseline model Fig. 2.

[c]Computed $E = B\sqrt{\omega/\mu_0 \sigma_a}$.

[d]20-m baseline.

[e]Joint variance = sum of variances.

[f]Signal integration 9–63 min (see Table 4) assuming 30 cycles at any frequency are required to achieve raw SNR.

**Table 3**. Relevant Environment Parameters for Europa

| Parameter | Value | Notes |
|---|---|---|
| **EMS-1** | | |
| Pressure | ≈0 | Vacuum |
| Temperature | 100 to 130 K (Moore et al., 2009) | Tropical-to-midlatitude daytime |
| Vault Temperature | -40 to +50°C (Krajewski et al., 2018) | Contains EMS electronics |
| Magnetic Field | 460 nT (Kivelson et al., 2004) | Magnetometer dynamic range |
| Surface Gravity | 1.6 ms$^{-2}$ | Electrode launcher performance |
| **EMS-2** | | |
| Electron Density | 10$^4$ cm$^{-3}$ (Kliore et al., 1997) | Electrometer performance |
| Electron Temperature | 20 eV (Saur et al., 1998) | Electrometer performance |
| **EMS-Flight** | | |
| Total Ionizing Radiation Dose | 1.7 Mrad (Krajewski et al., 2018) | Behind 100 mil Al (Si equivalent) |
| Total Ionizing Dose in Vault | 300 krad (Krajewski et al., 2018) | Si equivalent, includes RDF 2 |

**Table 4**. EMS Resource Requirements

| Parameter | Current Best Estimate | Maximum | Notes |
|---|---|---|---|
| Mass, g | 3510 | 4900 | Incl. 820-g titanium enclosure for internal electronics |
| Volume, L | | | |
|     Internal Electronics | 2.2 | 3.1 | |
|     External Sensors (Stowed) | 3.6 | 4.7 | 3 electrodes, 1 magnetometer |
| Power, W | | | |
|     Deployment | 15 | 20 | 4 mechanisms @ 3 min ea. |
|     Operation | 6.1 | 8.5 | |
|     Sleep | 0.1 | 0.4 | |
| Energy, W-hr | 120 | 170 | |
| Data, Mbit | 90 | 100 | Compressed |

**Table 5**. EMS Preliminary Concept of Operations (Conops)

| ELST, hr | 0–6 | 6–7 | 7–10 | 10–11 | 11–13 | 13–14 | 14–17 | 17–18 | 18–24 |
|---|---|---|---|---|---|---|---|---|---|
| Lander | Sleep | Sleep | Sample[a] | Sleep | Sleep | Sleep | Sample | Sleep | Sleep |
| EMS | Sleep[b] | Active | Off[c] | Active | Sleep | Active | Off | Active | Sleep |
| Data Acquisition Detail[d] | | | | | | | | | |
| Sampling Rate, Hz | | | 1 | 32 | 256 | | | | |
| Duration, min | | | 70 | 40 | 9 | | | | |

[a]Not operated during lander activity due to potential noise.

[b]Daytime operation only.

[c]EMS turned on just before lander sleep period.

[d]Allows 1 min boot.

Figures

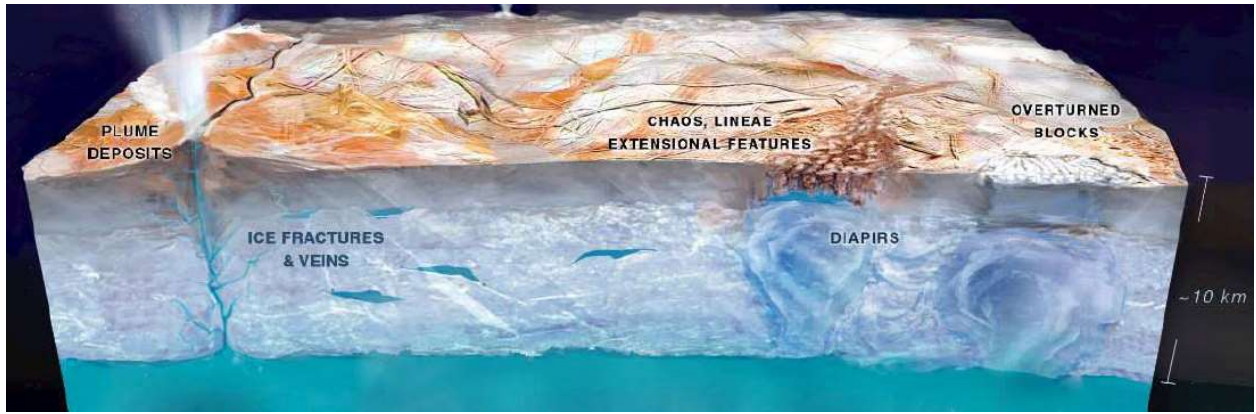

**Fig. 1.** Conceptual structures in the Europa ice shell (*Hand et al.*, 2016, used by permission) include subvertical dikes, subhorizontal sills ("ice fractures and veins") and quasi-spherical diapirs. Objective to "search for any subsurface liquid water within 30 km of the lander, including the ocean" is optimally addressed using low-frequency electromagnetics, specifically the magnetotelluric method. Note, the thickness of the Europan ice shell is estimated at 10 km, but could vary between 3-30 km thick.

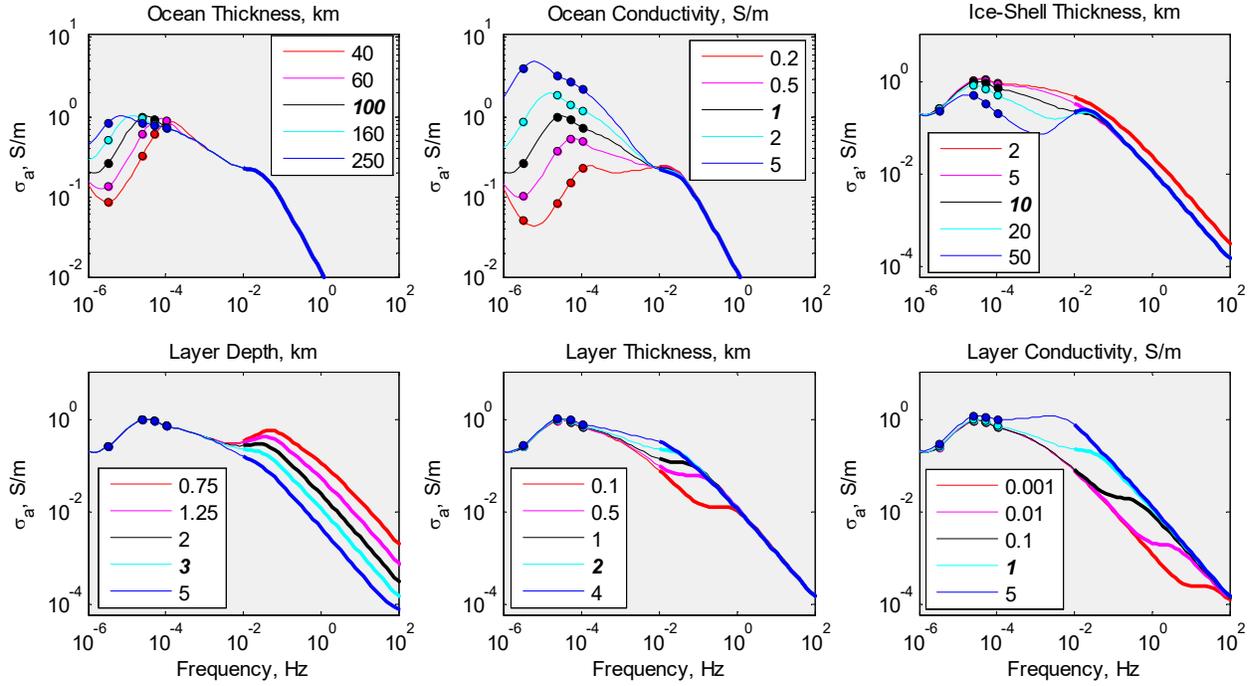

**Fig. 2**: Parametric study of EM sounding of Europa with result expressed as apparent conductivity $\sigma_a$. Reference value for each parameter (i.e., that used in other panels) is labeled in bold italic. Symbols are discrete "kinematic" frequencies (e.g., synodic period); heavy lines specify MT range to measure broadband magnetospheric phenomena. Kinematic frequencies can determine ocean properties and shell thickness, but magnetospheric frequencies are necessary to resolve water-layer properties. See text for details.

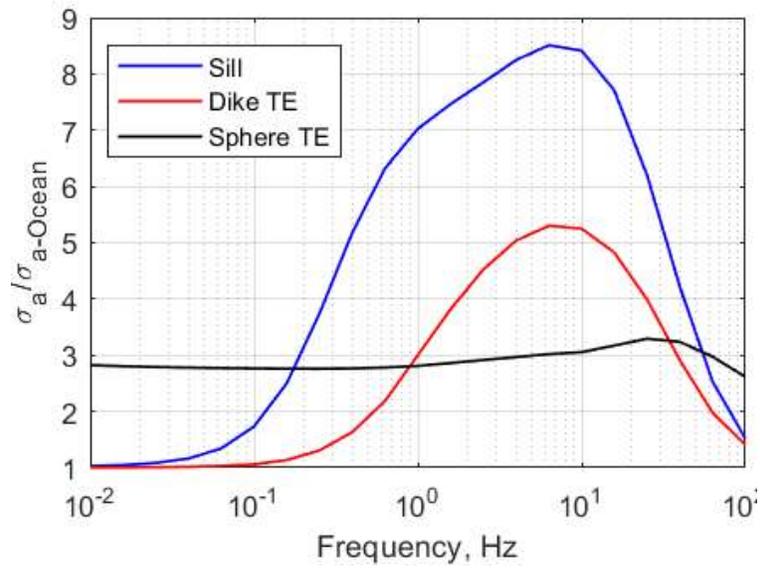

**Fig. 3.** 3D calculations of normalized apparent conductivity for three possible intrashell water-body geometries, expressed relative to response of the underlying ocean. By measuring both horizontal components of both electric and magnetic fields, the transverse electric (TE) mode can provide some basic shape discrimination. See text for details.

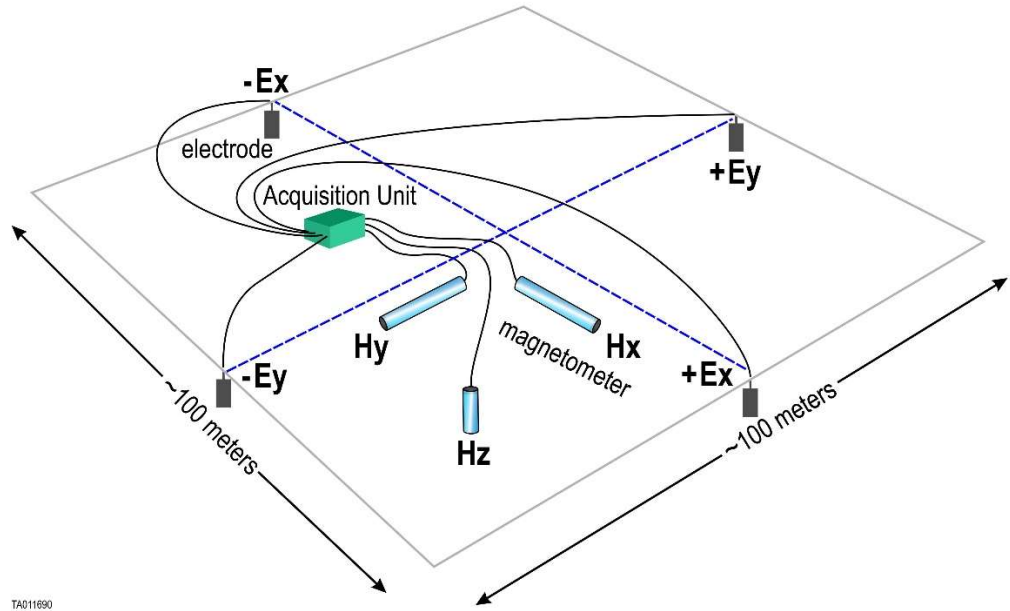

**Fig. 4.** Schematic terrestrial MT setup, using three large, single-component searchcoil magnetometers and porous-pot electrodes on orthogonal 100-m baselines (after Moombarriga Geoscience).

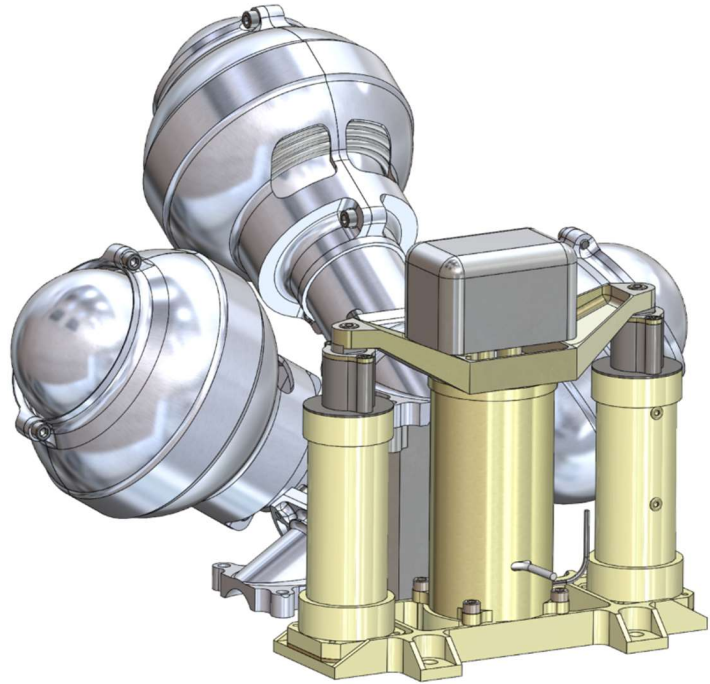

**Fig. 5.** EMS Magnetometer mast (gold) and sensor (gray), and three electrodes in launchers (silver) in notional compact configuration. 18 cm L x 19 cm W x 15 cm H. Alternatively, electrode launchers can be distributed around lander for larger baselines and higher signal.

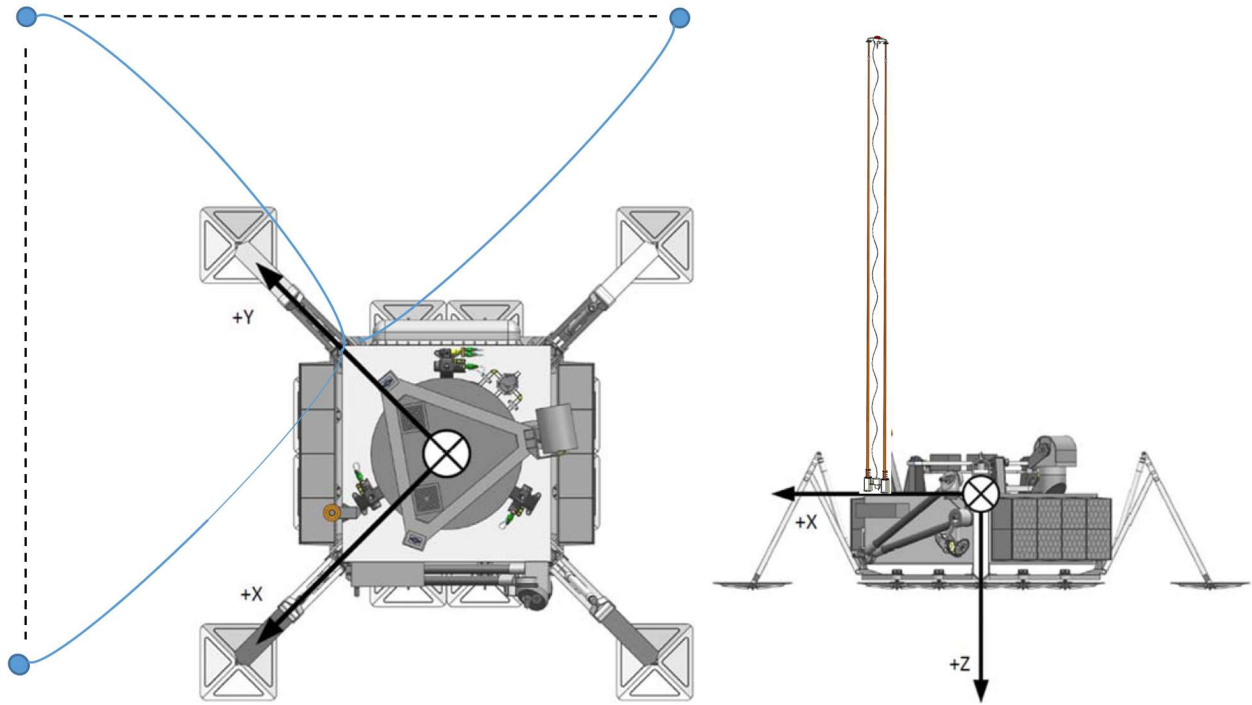

**Figure 6**. Schematic illustration of EMS on Europa Lander (Krajewski et al.,2018), as proposed. Sensors are deployed from +Y corner. <u>Left</u>: Electrodes are launched to 14 m distance (2 m illustrated); 90° separation yields orthogonal 20-m baselines (dotted) for electric-field measurement. <u>Right</u>: 2.5-m dual-Stacer mast (2 m illustrated) provides magnetometer standoff from Lander with accurate orientation knowledge.

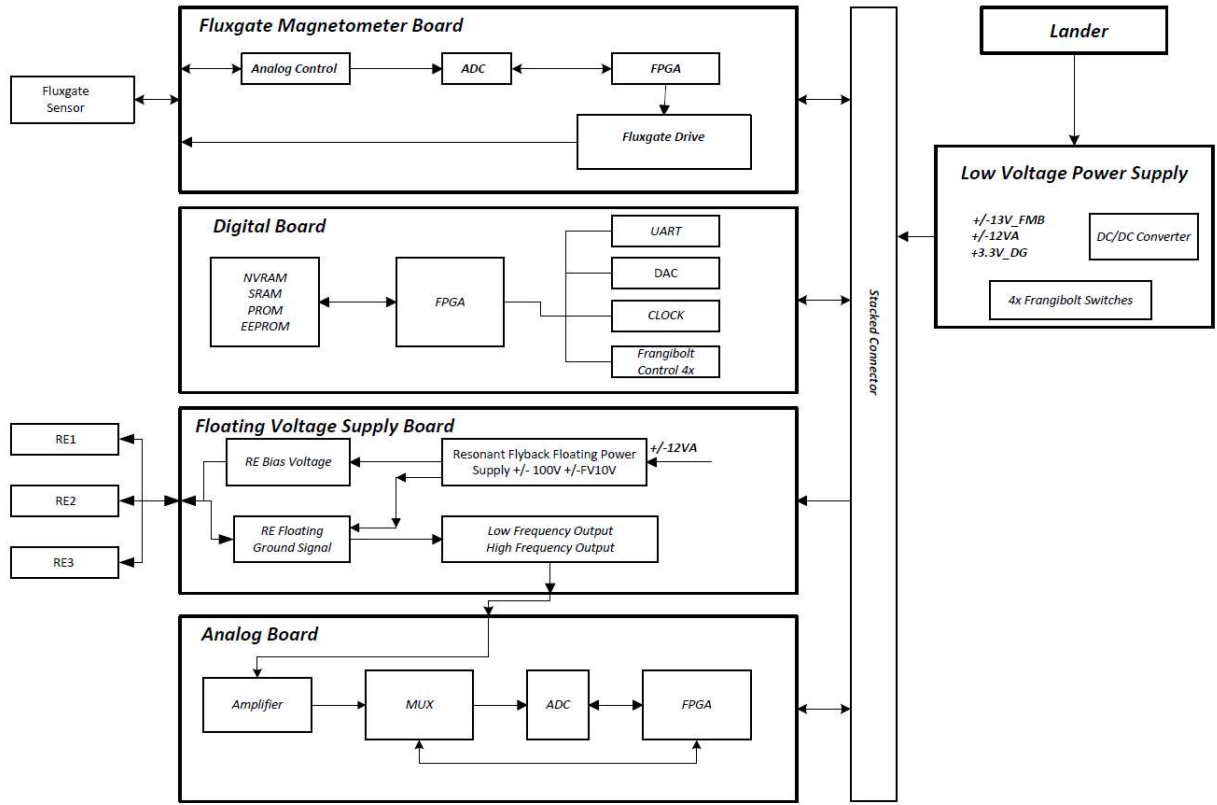

**Fig. 7.** Simplified block diagram of EMS-2.

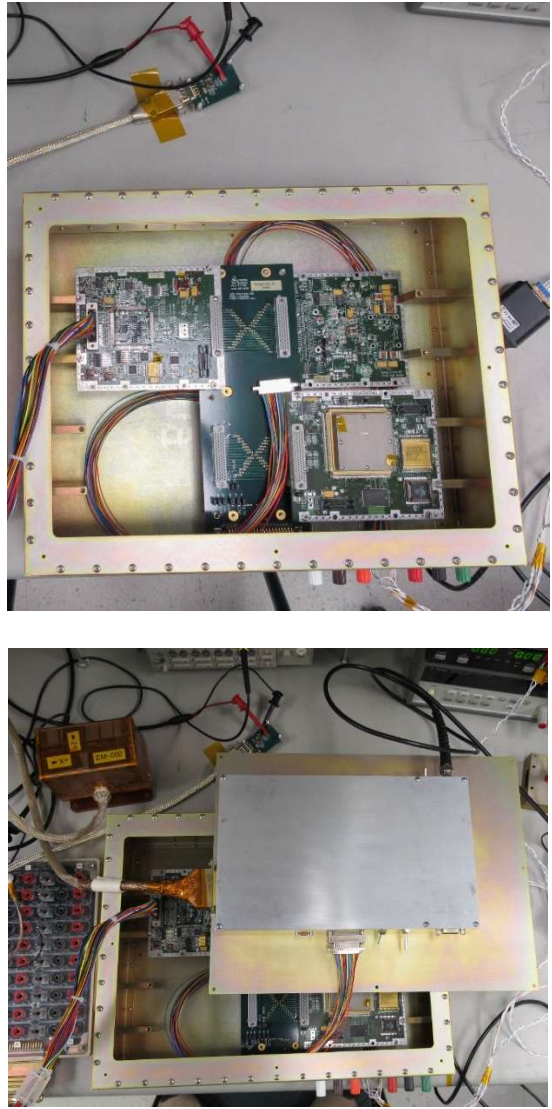

**Fig. 8.** Assembled electronics for EMS-1. <u>Top</u>: Flatsat enclosure with (l. to r.) AB, backplane, LVPS, DB. FVSB not included in this build. Note remote-electrode circuit board at top with test input. <u>Bottom</u>: With MB in ground-support enclosure and kapton-wrapped MAVEN fluxgate sensor upper left.

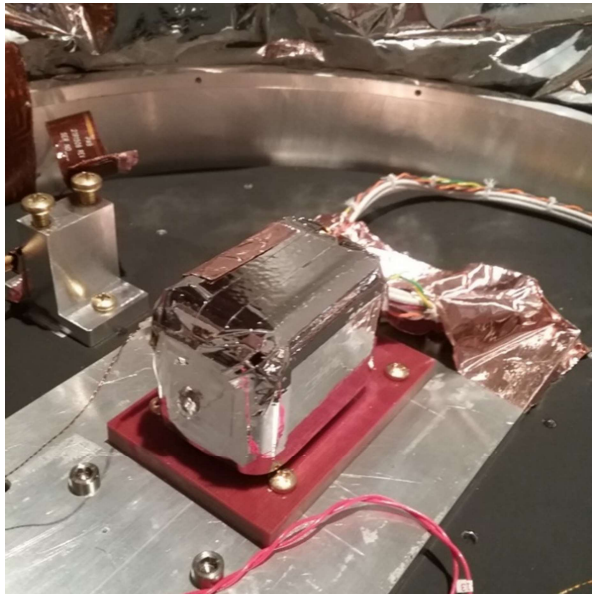

**Fig. 9.** POGS fluxgate sensor in the non-magnetic thermal vacuum chamber undergoing testing at 80 K, through viewport.

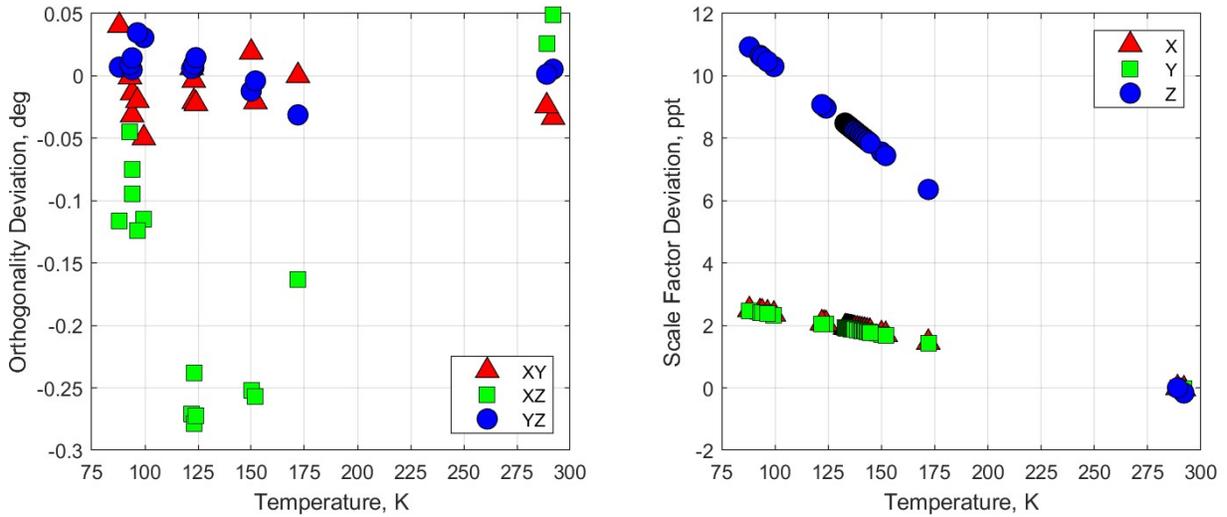

**Fig. 10.** Fluxgate magnetometer performance as a function of temperature. <u>Left</u>: Orthogonality deviation (difference from 90°). XY and YZ orthogonality are always within ±0.05° (0.6‰) with XZ maximum excursion <0.3° (3.3‰). <u>Right</u>: Scale factor deviation is small and linear and so allows good calibration to levels <1‰. These results indicate that the magnetometer can operate unheated at Europa with 15 pT relative accuracy or better with respect to the relevant source fields ~3 nT.

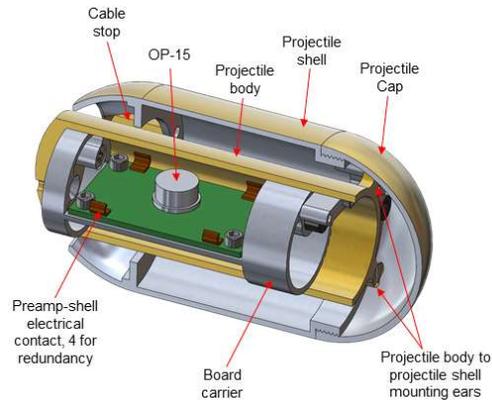

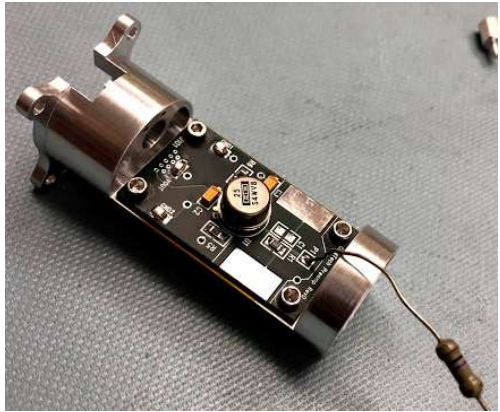

**Fig. 11**. <u>Top</u>: Sensing electrode. Cable trails from left. <u>Bottom</u>: Assembled electrode chassis and electronics. Mounting differs slightly from drawing.

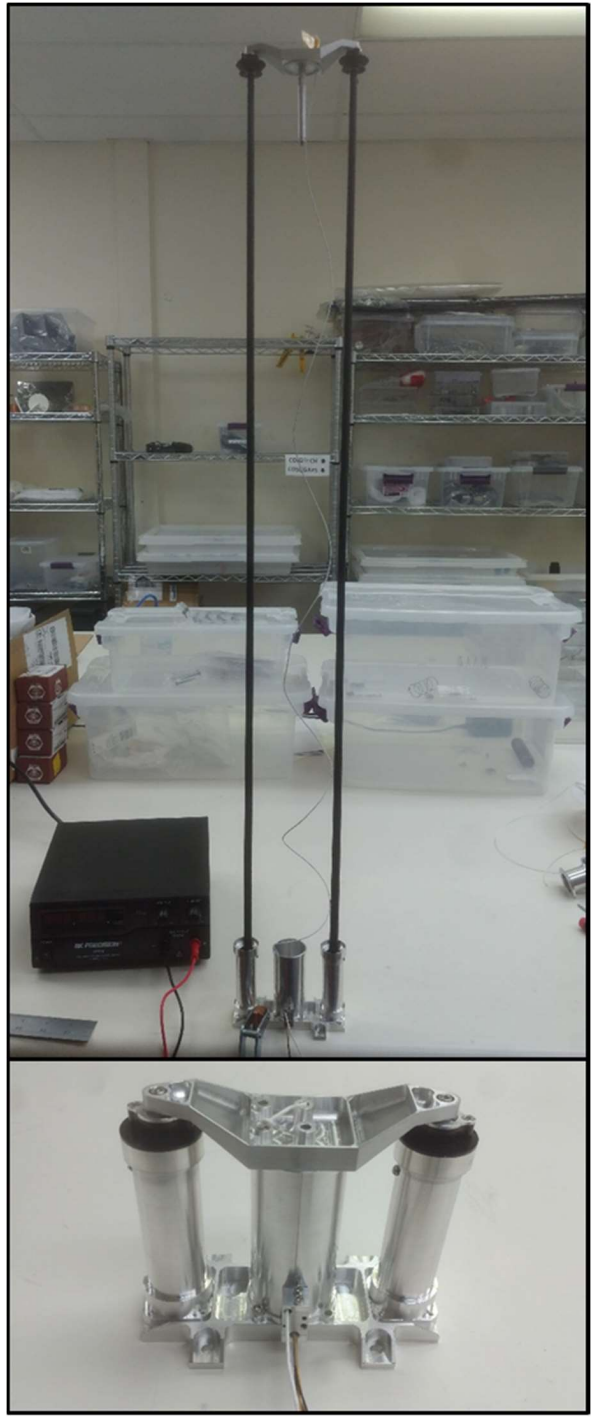

**Fig. 12.** Magnetometer mast. <u>Top</u>: Deployed 2-m protoype; final mast 2.5 m. <u>Bottom</u>: Stowed prototype. POGS-sized fluxgate sensor fits on top. Stowed dimensions with sensor 12 x 6 x 12 cm.

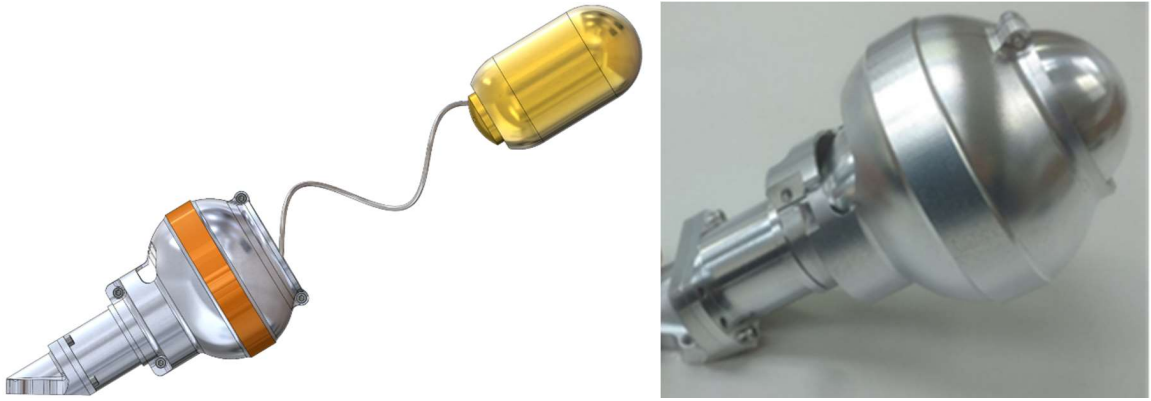

**Fig. 13.** Electrode deployment. <u>Left</u>: Frangibolt releases spring, which ejects electrode. Cable is wound inside widest part of launcher. Right: Assembled launcher with stowed electrode. Stowed dimensions 15 x 8 cm.

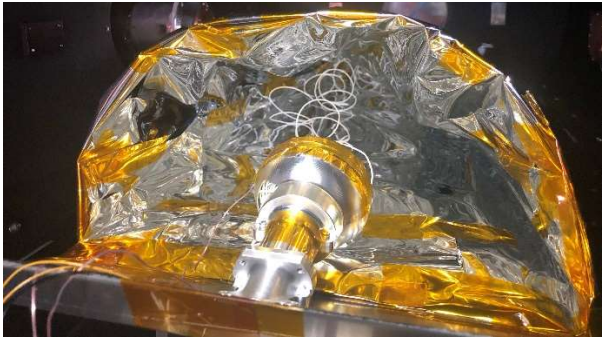

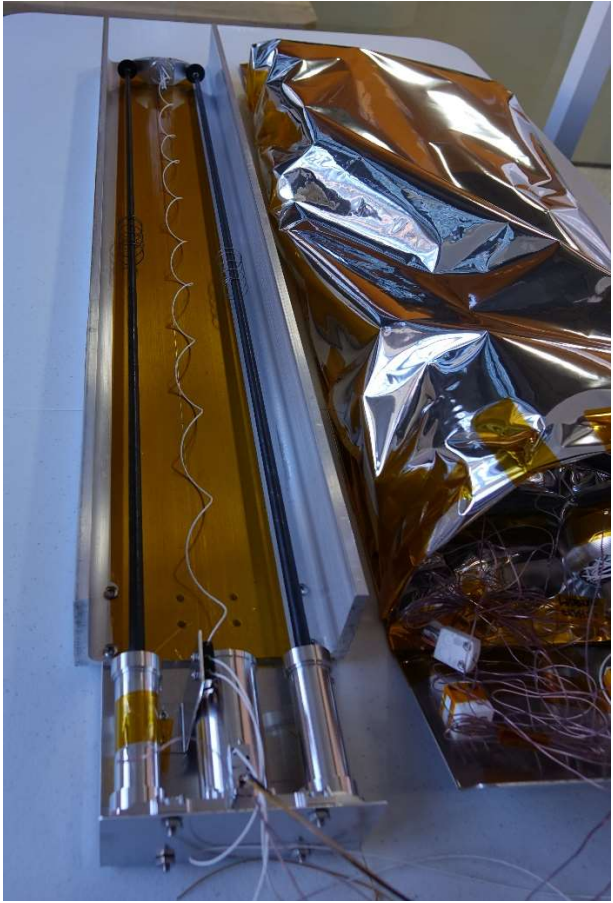

**Fig. 14.** TVAC deployment tests. <u>Top</u>: Electrode launcher, still in chamber. <u>Bottom</u>: Magnetometer mast (with electrode catcher net rolled up) removed from chamber. Connecting cables are multiconductor prototypes.

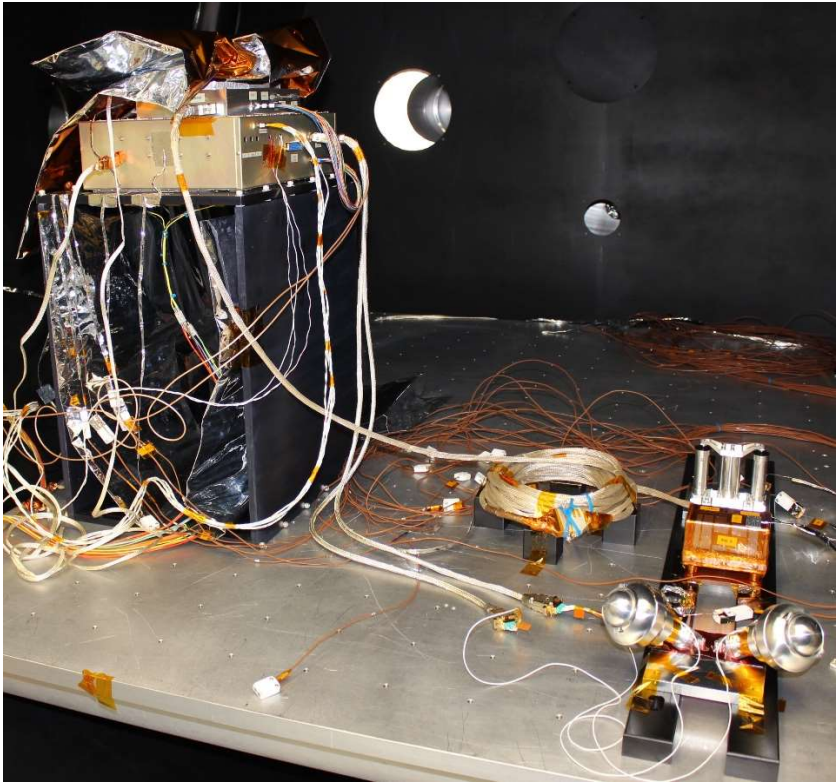

**Fig. 15**. EMS in thermal-vacuum chamber for functional test, prior to door closing. Baseplate and walls were set to Europa temperature, with sensors and mechanisms (right) thermally coupled. Deployment mechanisms were not activated in this electronics functional test and oversized fluxgate sensor is not attached to mast. Elevated and blanketed electronics boxes (left) were maintained at vault temperature.

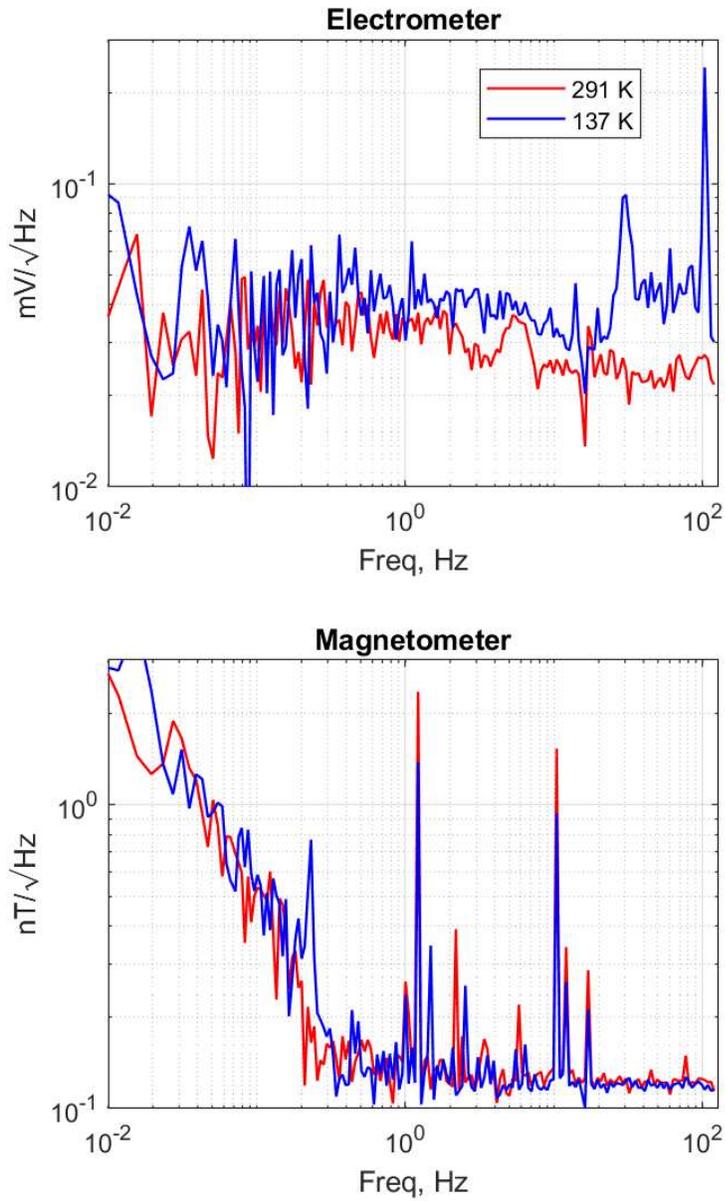

**Fig. 16.** Functional thermal-vacuum test. Powerline noise is removed in both sensors; remaining narrowband signals in magnetic field are likely microphonics from facility vibration. Noise in voltage measurement at higher frequencies is unexplained.

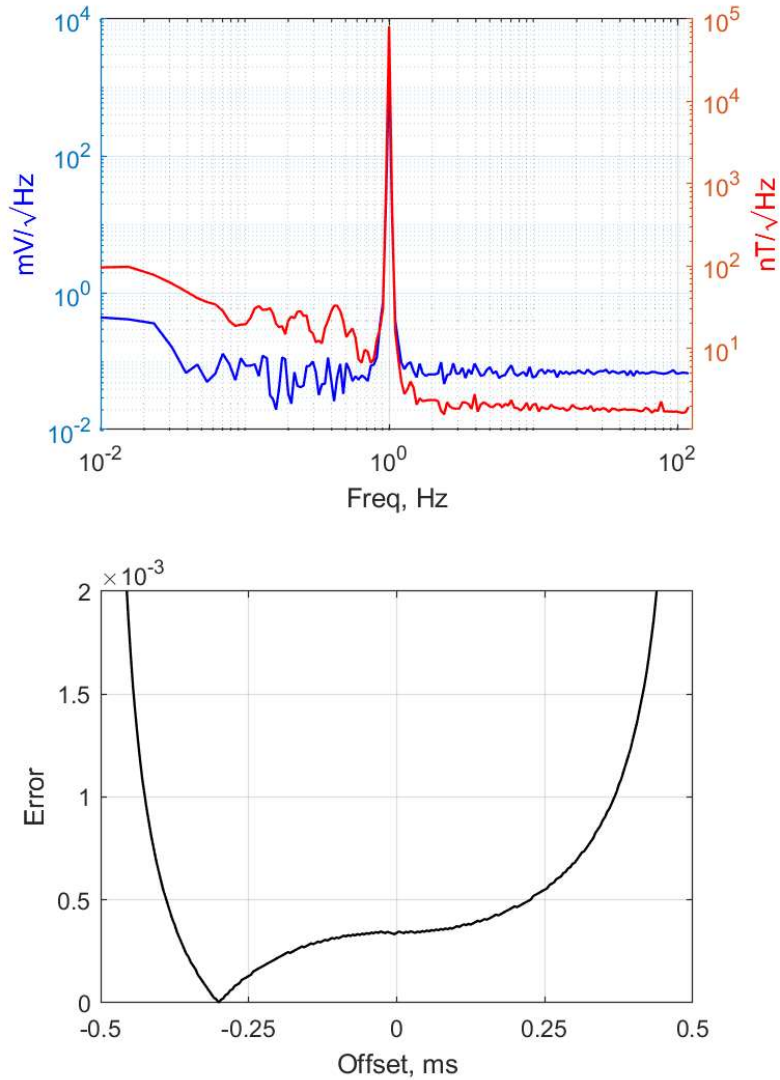

**Fig. 17.** Room-temperature signal-offset test. <u>Top</u>: Spectra of input signals. A voltage difference was applied jumpered across the electrodes and a coil wrapped around the magnetometer in a mu-metal can. <u>Bottom</u>: Correlation error of interpolated time series as a function of offset. Best fit at −0.28 ms is much less than one sample (3.9 ms).

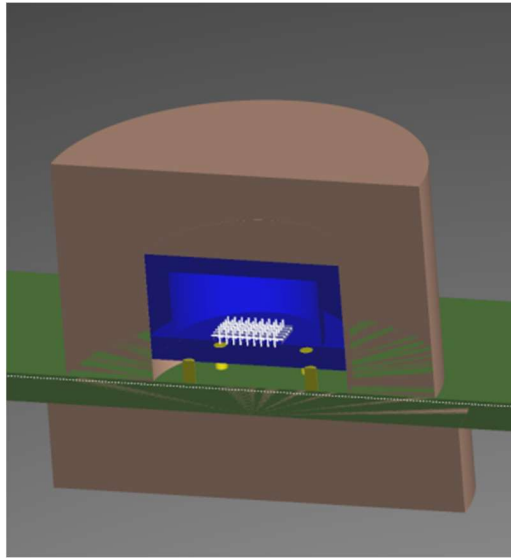

**Fig. 18.** Radiation mitigation design for remote electrodes. Tantalum shielding (tan) is split into parts above and below PCB (green) and surrounds TO-99 can (blue) containing remote-electrode op-amp. Split design minimizes shielding volume (mass) at only small loss in efficiency. Upper shield is 16 mm diameter.